\documentstyle[12pt,epsfig]{article}
\begin{document}
\def\bfg #1{{\mbox{\boldmath $#1$}}}
\begin{center}
{\large \bf { REACTION $pp\to \{pp\}_s\pi^0$
 IN THE GeV REGION AND $\pi^0 p$ RESCATTERING}}
\vskip 1em
 Yu.N. Uzikov
\footnote{e-mail address: uzikov@nusun.jinr.ru\\}

{\it Laboratory of Nuclear Problems, Joint Institute for Nuclear Research\\
 Dubna, Moscow reg., 141980 Russia}
\end{center}
\vspace{0.3cm}

{\small {\bf Abstract}
\vspace{0.3cm}

  COSY data on the  cross section of the reaction $pp\to \{pp\}_s\pi^0$,
 where $\{pp\}_s$ is the proton pair in the $^1S_0$ state 
 at small excitation energy $E_{pp}=0-3$ MeV,
 recently obtained for  beam energies 0.5 - 2.0 GeV
 are analyzed within the one-pion exchange model. The model is based 
 on the subprocess $\pi^0 p\to \pi^0 p$ and final state pp-interaction.
   A broad maximum observed in the energy dependence of the cross section
 at 0.5 - 1.4 GeV  in the forward direction is explained by this model
 as a dominant contribution of the isospin $\frac{3}{2}$ in 
 the $\pi^0 p$-rescattering.
 The second maximum observed  at 2 GeV is
 underpredicted within the model
 by one order of magnitude.  
}

\vspace{0.3cm}

{\it Keywords}: Pion production; Final state interactions\\
PACS: 13.75.Cs; 13.60.Le; 25.40.Qa

   \section{Introduction}

  Study of the reaction
  $pp\to \{pp\}_s\pi^0$, where  $\{pp\}_s$ is the
  proton pair (diproton) in  the $^1S_0$ state at small
 excitation energy $E_{pp}=0-3$ MeV,
 is motivated by several reasons.
 First, this is the
 simplest inelastic process in the pp-collision, which can reveal
underlying dynamics of NN interaction.
Second, restriction to only one pp-partial wave
(s-wave) in the final state
 considerably simplifies a comparison with theory making
 it basically similar to that for the other simplest 
reaction of this type,  $pp\to d\pi^+$. 
  However, while for the reaction
 $pp\to d\pi^+$ there are a lot of data including spin observables 
\cite{arndt}, which are used to test
  theoretical models in the GeV region \cite{GarcilazoM,hanhart}, data on the 
 reaction $pp\to \{pp\}_s\pi^0$ above 0.4 GeV were absent until 
 recent measurements  at  COSY \cite{dymov06,kurbatov}.
  Third, the quasi-binary reaction $pp\to \{pp\}_s\pi^0$
  is very similar kinematically to the reaction $pp\to d\pi^+$, but its
 dynamics can be essentially different. In fact, quantum numbers of the
 diproton state ($J^\pi=0^+,\,I=1,\, S=0, \, L=0$) differ from these for
 the deuteron ($J^\pi=0^+, I=0,\, S=1, L=0,2$). Therefore, transition
 matrix elements for these two reactions are also different.
  Using the generalized Pauli principle and
   angular momentum and P-pariry conservation, one can  easily find that
 only negative parity states  are allowed in the reaction
 $pp\to \{pp\}_s\pi^0$. Thus, for the intermediate $\Delta N$ state 
 odd partial 
waves (p-, f-, $\dots$) are allowed, whereas  even waves (s-, d-, $\dots$)
 are forbidden. Therefore, at the nominal $\Delta(1232)$-threshold 
 of the reaction $ NN\to \Delta N$, $T_p=0.63$ GeV,
 the lowest allowed partial wave is the p-wave, which, however, 
 has to be suppressed by the centrifugal barrier. In contrast, in the 
$pp\to d\pi^+$ reaction both negative and positive parity $\Delta-N$ states 
are allowed. As a consequence, the relative contribution of the 
$\Delta$-mechanism to the reaction $pp\to \{pp\}_s\pi^0$ is expected to be
suppressed as compared to the reaction $pp\to d\pi^+$. This argument was
 applied in Ref. \cite{uzwilk2001} to explain   a very 
 small ratio 
(less of few percents) of the spin-singlet to spin-triplet
 pn-pairs observed in the LAMPF data \cite{HGabitch}
 in the final state interaction region of the reaction
 $pp\to pn\pi^+$ at proton beam energy 0.8 GeV. 
 Obviously, this
 argument is valid for any intermediate $N^*N$- states with other nucleon
isobars $N^*$  of positive parity. Furthermore, since $\Delta-$type 
mechanisms are of long-range type,
 reduction of their contribution would mean that other mechanisms, like
$N^*$-exchanges \cite{sharmamitra} which are more sensitive to short-range
 NN-dynamics, could be more important in the reaction
 $pp\to \{pp\}_s\pi^0$ as compared to 
the $pp\to d\pi^+$ reaction \cite{ponting}.

  The cross section of the reaction $pp\to \{pp\}_s\pi^0$ was measured recently
  at energy  0.8 GeV in Ref.\cite{dymov06} and  
  at beam energies 0.5 - 2.0 GeV in Ref. \cite{kurbatov}.
  (For measurements at energies below 0.425 GeV
  see Refs.\cite{meyer, maeda, bilger}.)
  At the zero angle, the data \cite{kurbatov} show 
  a broad maximum in the energy dependence of
  the cross section at 0.5 -1.4 GeV. This maximum is similar in shape and 
  position to the well known $\Delta-$ maximum in the reaction $pp\to d\pi^+$.  
  However, a comparison with the   calculation   \cite{niskanen06} performed 
  within a microscopical  model, which  includes
 $\Delta(1232)$-isobar  excitation and s-wave $\pi N$-rescattering,
 shows  very strong disagreement between the model and the data obtained  at
 energies 0.5 - 0.9 GeV \cite{kurbatov} both in the absolute value
 and shape of energy dependence  of the cross section. So, the forward
 cross section measured in Ref. \cite{kurbatov} is lower than the calculated 
 one \cite{niskanen06}
  by factor of three at 0.6 GeV, where the $\Delta-$isobar
 maximum would be  expected, whereas at 0.5 and 0.8 GeV the disagreement is 
 more than one order of magnitude \cite{kurbatov}.    

  \begin{figure}[hbt]
\mbox{\epsfig{figure=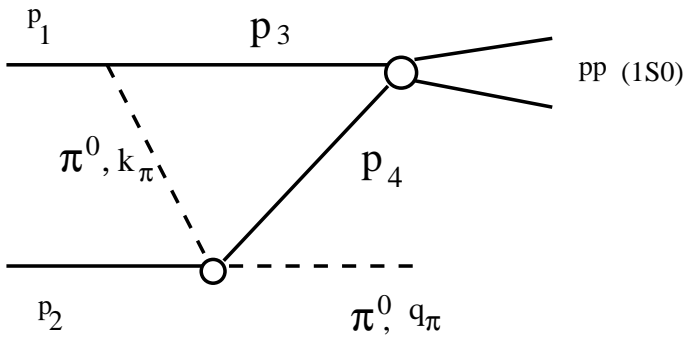,height=0.2\textheight, clip=}}
\caption{The OPE mechanism of the reaction $pp\to \{pp\}_s\pi^0$. }
\label{fig1}
 \end{figure}
   In view of qualitative arguments given above, this disagreement
 would mean that either the model of Ref. \cite{niskanen06} is incorrect,
 or the observed maximum of the cross section of the
 reaction $pp\to
   \{pp\}_s\pi^0$ at 0.5 - 1.4 GeV is of non-$\Delta$-isobar origin
\footnote{It was supposed in Ref.\cite{kurbatov} 
  that the observed maximum is caused by the $\Delta-$isobar contribution,
  but no theoretical calculation were performed to confirm it.}.
  Here we analyse these data employing a simpler model,
  which includes the subprocess $\pi ^0 p\to \pi^0 p$ and the final
  state pp$(^1S_0)$-interaction (Fig.\ref{fig1}). 
  We show that the observed shape of the peak
  and, to a large extent, its magnitude are in agreement with the dominance of
  the $\Delta(1232)$-isobar contribution.

\subsection{ The model}

  We consider the reaction $pp\to \{pp\}_s\pi^0$ within the mechanism
 which corresponds to the triangle diagram
 in Fig. \ref{fig1}.  A very similar mechanism was
 successfully
 applied for analysis of the $pp\to d\pi^+$ reaction
in the region of
 the $\Delta(1232)$-isobar \cite{yao,locher} and at higher energies
 too \cite{yao}.

 The amplitude of the reaction $pp\to \{pp\}_s\pi^0$
 consists of two terms, $A=A^{dir}-A^{exch}$,
 where $A^{dir}$ is the direct term and $A^{exch}$  is the exchange one.
 These terms are related to one another  by permutation 
 of  two initial protons. The one-loop integral for  the direct term $A^{dir}$
 is  evaluated very similarly to 
 the OPE-II model of the reaction $pd\to \{pp\}_sn$  considered in
 Ref.\cite{uzhw2007}. Thus, $A^{dir}$ takes the following form:
\begin{eqnarray}
\label{direct}
A^{dir}(p_1,\sigma_1,p_2,\sigma_2)
=\frac{f_{\pi NN}}{m_{\pi}}N_{pp}2m_pF_{\pi NN}(k^2_{\pi})\times
\\ \nonumber
\times \Sigma_{\sigma_3\, \sigma_4\, \mu}
(\frac{1}{2}\sigma_3\frac{1}{2}\sigma_4|00)
(1\mu\frac{1}{2}\sigma_3|\frac{1}{2}\sigma_1)J^{\mu}({\tilde p},\gamma)
A_{\sigma_2}^{\sigma_4}(\pi^0p\to \pi^0p),
\end{eqnarray}
here $f_{\pi NN}$ is the $\pi NN$ coupling constant with 
${f^2}_{\pi NN}/4\pi=0.0796$, $F_{\pi NN}({k^2}_\pi)=(\Lambda^2-{m^2}_\pi)/
(\Lambda^2-{k^2}_\pi)$ is the $\pi NN$ form factor, $k_\pi$ is the
four-momentum of the virtual $\pi$-meson,
$m_p$ ($m_{\pi}$) is the nucleon (pion) mass, $\sigma_i$ ($i=1,\dots,4$) is
the z-projection of the spin of {\it i}th proton; $A(\pi^0 p\to \pi^0 p)$ is
the amplitude of the $\pi^0 p$ elastic scattering which is taken on-mass-shell; 
the vector  $J_{\mu}$ is defined by the transition form factors  as
\begin{eqnarray}
\label{jmu}
J^{\mu}({\tilde p},\gamma)=\sqrt{\frac{E_1+m_p}{2m_p}}\frac{m_p}{E_1}\Bigl\{
 R^{\mu}F_0({\tilde p},\gamma)-i{{\hat {\tilde  p}}}^{\,\mu}
\Phi_{10}({\tilde p},\gamma)\Bigr\},
\\ \nonumber
\end{eqnarray}
where
\begin{eqnarray}
\label{fphi02}
F_0({\tilde p},\gamma)=\int_0^\infty dr r j_0({\tilde p}r)\psi_{k}^{{(-)^*}}(r)
\exp{(-\gamma r)},\\
\label{phi02}
\Phi_{10}({\tilde p},\gamma)= i\int_0^\infty dr j_1({\tilde p}r)
\psi_{k}^{{(-)}^*}(r)(1+\gamma r)
\exp{(-\gamma r )},
\end{eqnarray}
 here $j_l(x)$ ($l=0,\,1)$ is the spherical Bessel function,
 $\psi_{\bf k}^{(-)}({\bf r})$ is the  pp-scattering wave function
 that is the solution of the
 Schr\"odinger
 equation  at the cms 
 momentum $|{\bf k}|$ 
 with  the interaction potential $V(^1S_0)$ for
 the following boundary condition  at $r\to \infty$:
\begin{equation}
\label{boundary}
\psi_{\bf k}^{(-)}({\bf r})\to \frac{sin(kr+\delta)}{kr}. \,
\end{equation}
Here $\delta$ is the $^1S_0$ phase shift (for simplicity
we omit here the Coulomb interaction, which is taken into account
in real  numerical calculations).
In Eq.(\ref{direct}) the combinatorial factor $N_{pp}=2$  
 takes into account identity of two protons.
Kinematical variables in Eqs. (\ref{jmu}) - (\ref{phi02}) 
are defined as
\begin{eqnarray}
\label{kinem2}
\gamma^2=\frac{T_1^2}{(E_1/m_p)^2}+\frac{m^2_{\pi}}{E_1/m_p},\ \ 
{\bf R}=-{\bf p}_1\frac{m_p\,T_1}{(E_1+m_p)E_1},\ \
{\bf {\tilde  p}}=\frac{{\bf p}_1}{E_1/m_p},
\end{eqnarray}
where $E_1$, ${\bf p}_1$ and $T_1=E_1-m_p$ are the total energy,
 3-momentum and
 kinetic energy of the initial proton $p_1$, respectively, 
in the rest frame of the final diproton.
The exchange amplitude $A^{exch}$ can be obtained from
 Eqs.(\ref{direct})-(\ref{kinem2})
by interchanging $1\leftrightarrow 2$.

 The OPE cross section of the reaction $pp\to \{pp\}_s\pi^0$ in the cm system 
is
\begin{eqnarray}
\label{unpolarized}
\frac{d\sigma}{d\Omega}=\frac{1}{(4\pi)^5}\frac{p_f}{s_{pp}\,p_i}
\int_0^{k^{max}}dk^2 \frac{k}{\sqrt{m^2_p+k^2}}\frac{1}{2}
\int d\Omega_{\bf k}
{\overline {|A_{fi}|^2}},
\end{eqnarray}
where $k^{max}$ is the maximal relative  momentum in the final pp-system,
related to the maximal relative energy ${E^{max}}$ as 
$k^{max}=\sqrt{{E^{max}}m_p}$,
$p_i$ $(p_f)$ is the cms momentum in the initial (final) state of the $pp\to
\{pp\}_s\pi^0$ reaction, $s_{pp}$ is the squared invariant mass of the initial
 pp-system. The factor $\frac{1}{2}$ in front of the integral over directions
 of ${\bf k}$ takes into account identity of two final protons.
Keeping only the direct term of Eq.(\ref{direct}), one can  finally find   
 from Eq. (\ref{unpolarized})
%
 \begin{eqnarray}
\label{pppppi0}
\frac{d\sigma}{d\Omega}_\theta (pp\to \{pp\}_s\pi^0)=
\frac{1}{24\pi^2}\frac{p_f}{p_i}\frac{s_{\pi p}}{s_{pp}}
\Biggl [\frac{f_{\pi NN}}{m_{\pi}}N_{pp}m_pF_{\pi NN}(k_{\pi}^2)\Biggr]^2\times
\nonumber\\ 
\times \int_0^{k^{max}} dk \frac{2k^2}{\sqrt{m_p^2+k^2}}
\Bigl \{ 2|J^{\mu=0}({\tilde p},\delta)|^2+|J^{\mu=1}({\tilde p},\delta)|^2
\Bigr\}\frac{d\sigma}{d\Omega}_\phi (\pi^0 p\to \pi^0 p).
\end{eqnarray}
The differential crosss section  of the reaction $\pi^0 p\to \pi^0 p$ is taken  
in Eq. (\ref{pppppi0})  at the squared invariant mass of the $\pi p$ system,
 $s_{\pi p}$,
defined as
\begin{eqnarray}
\label{sxp}
s_{\pi p}=(m_\pi+ m_p)^2+2T_\pi m_p.
\end{eqnarray}
Here $T_\pi$ is the kinetic energy of the final meson $\pi^0$ in the rest frame
of the final diproton.
If $\theta$ is the  angle between  the cms momenta of the diproton 
and the proton $p_1$, which  emits the virtual pion in the direct
 OPE diagram in Fig.\ref{fig1}, and $\phi$ is the cms scattering angle of
 the $\pi^0$-meson  in  the process 
$\pi^0(k_{\pi})+ p_2\to p_4+\pi^0(q_\pi)$, then  one can find the following
relation: 
\begin{eqnarray}
\label{cmangles}
p_{20}q_{0}+|{\bf p}_2||{\bf q}_\pi|\cos{\phi}=
\sqrt{m_p^2+{ p}_i^2}\sqrt{m_{\pi}^2+{ p}_f^2}-p_i p_f\cos{\theta},
\end{eqnarray}
where the four-momenta of the initial proton $p_2=(p_{20},{\bf p}_2)$
and the final  $\pi^0$-meson  $q_\pi=(q_{0},{\bf q}_\pi)$ in the cms of the
 $\pi p$ system
  can be written as
\begin{eqnarray}
\label{kinsubprocess} 
p_{20}=\frac{1}{2\sqrt{s_{\pi p }}}(s_{p\pi}+m_p^2-k_{\pi}^2),\,\,
q_{0}=\frac{1}{2\sqrt{s_{\pi p }}}(s_{\pi p}+m_{\pi}^2-m_p^2),\nonumber\\ 
|{\bf q}_\pi|=\sqrt{q_{0}^2-m_\pi^2}, \,\,
|{\bf p}_2|=\sqrt{p_{20}^2-m_p^2}.
\end{eqnarray}
The squared four-momentum  of the intermediate $\pi$-meson is
\begin{eqnarray} 
\label{kpi2}
k_{\pi}^2=2m_p^2+p_i p_f
\cos{\theta}-\sqrt{m_p^2+p_i^2}\sqrt{M_{pp}^2+p_f^2},
\end{eqnarray}
where $M_{pp}$ is the mass of the final diproton.
One can find from Eqs. (\ref{cmangles}), (\ref{kinsubprocess}) and
(\ref{kpi2}) 
that backward $\pi^0 p$ scattering ($\phi=180^\circ$) dominates diproton
formation in the forward direction ($\theta=0^\circ$).
   
 Analysis   of the reaction $pp\to d\pi^+$  in the $\Delta$ region, performed
 in Refs.  \cite{locher} shows
 that the contribution of the pole diagram with the neutron
   exchange is small but non-negligible and being added to the OPE diagram
 with $\pi N$ rescattering
   improves the agreement with the data. For the reaction $pp\to
   \{pp\}_s\pi^0$ a similar pole diagram seems to be less important
   and is not taken into account here. The point is that at 
 $T_p=0.5-2.0$ GeV and $\theta=0^\circ$ 
    the $pp\to\{pp\}_s$ vertex in the pole diagram of the reaction  $pp\to
   \{pp\}_s\pi^0$ involves the high momentum component
 of the wave function  $\psi^{(-)}_{\bf k}({\bf q})$ at the relative momentum
 between protons $q=0.4-0.6$ GeV/c, but  in contrast to the $pn\to d$
 vertex, does not contain the D-wave which is important for the pole diagram 
 at large  $q$.
  Furthermore, the S-wave component has a node 
 at $q\approx 0.4$ GeV/c (see, for example, Ref.\cite{uzhw2007}).

\section{Numerical results and discussion}

\begin{figure}[hbt]

\mbox{\epsfig{figure=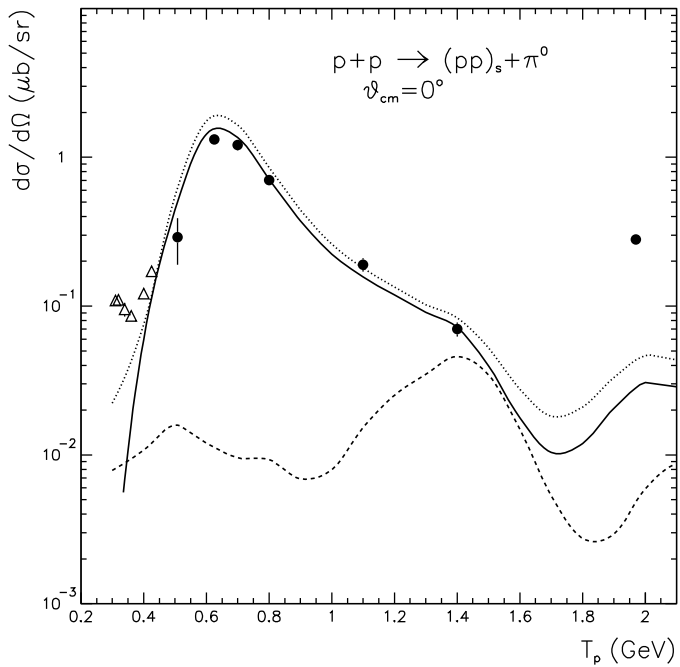,height=0.45\textheight, clip=}}
\caption{
The cms cross sections of the 
reaction $pp\to\{pp\}_s\pi^0$ at $\theta=0^\circ$
 versus beam energy.
 Data are taken from Ref. \cite{bilger}
($\triangle$) and \cite{kurbatov} ($\bullet$).
 The dotted curve presents the calculated cross section for 
 the incoherent sum of the direct and exchange
 terms of the OPE model amplitude.
 Other  curves 
 are obtained with the direct term $A^{dir}$ only
for the isospin term $a_{\frac{3}{2}}$ excluded from (dashed line) and included
(full line) in the $\pi^0p\to \pi^0p$ amplitude, as explained in the text.
 All curves are scaled by the factor $0.45$.
 }
\label{fig2}
\end{figure}
 In numerical calculation we used the data on the elementary $\pi N$ reactions
 from SAID \cite{arndt}. The scattering wave function $\psi^{(-)}_{\bf k}({\bf
 r})$ of the $pp$ system at low energy $<3$  MeV is largely independent of
 the NN model and is calculated here using the Reid soft core 
 potential plus Coulomb interaction \cite{reid}.   
 The calculated forward cross section 
 multiplied by the factor 0.45 is shown  in Fig. \ref{fig2}.
 When comparing the dotted and full lines in Fig. \ref{fig2}, one can see 
 that the contribution of the
 exchange term $|A^{exch}|^2$ is much less important
 than the direct term $|A^{dir}|^2$ at $\theta=0^\circ$.
 Thus, we neglect below the term $A^{exch}$.
 As seen from Fig. \ref{fig1},
 the OPE model is in good agreement
 with the observed shape of the cross section at 0.5 - 1.4 GeV.
 Note that the form factors $F_0({\tilde p,\gamma})$
 and $\Phi_{10}({\tilde p},\gamma)$ in Eq. (\ref{jmu}) are  smooth
 functions of the beam energy $T_p$. Therefore, the calculated shape of the 
 $pp\to \{pp\}_s\pi^0$ cross section follows mainly  the $T_p$-dependence
 of the $\pi^0 p\to \pi^0 p$ cross section at the cms angle $\phi=180^\circ$.  
 The  disagreement in absolute value by factor 0.45
 corresponds to  a typical factor of
 absorptive distortions in the initial $pp$ state \cite{hanhartnakayamahlee}.
 The distortions are not taken into account
 in the present work in view of their  dependence on  unknown 
 details of the production
 mechanism, in particular, on off-shell behaviour of the $\pi^0 p$-scattering
 amplitude. Furthermore, one should  note  that when calculating the diagram in
 Fig.\ref{fig1}, 
 we factor the amplitude of the elastic $\pi^0 p$ scattering
 outside   the  integral sign.
 Within this approximation,  
 the contribution of intermediate $\Delta-N$ states of
 positive parity  is not excluded from  this reaction as it should be
 in an exact OPE amplitude
 according to the discussion in the Introduction.
 For this reason,  this simple  model cannot  provide a precise
 absolute value of   
 the cross section of the reaction $pp\to \{pp\}_s\pi^0$.
 The disagreement in the absolute value of the cross
 section   can be also  related in part to the neglected off-shell effects 
 in the $\pi^0 p\to \pi^0 p$ amplitude
 and 
  contribution of the mesons $\eta$, $\eta'$ and $\omega$.

 In order to exhibit sensitivity of the calculated cross section 
 to the  $\Delta$-isobar contribution, one can
 completely exclude  the contribution of the isospin $\frac{3}{2}$
 from the $\pi^0 p$ scattering.
 The isospin decomposition of the $A(\pi^0 p\to \pi^0 p)$
 amplitude is the following:
\begin{equation}
\label{api0}
 A(\pi^0 p\to \pi^0 p)=\frac{1}{3}\Bigl ( a_{\frac{1}{2}}+
2a_{\frac{3}{2}}\Bigr ),
\end{equation}
 here $a_\frac{1}{2}$ ($a_\frac{3}{2}$)  is the amplitude with the total isospin
 $\frac{1}{2}$ ($\frac{3}{2}$).
 The cross section of the $\pi^0 p$ elastic scattering 
 can be written as
 \begin{equation}
\label{pi0psec}
d\sigma(\pi^0 p\to \pi^0 p)=\frac{1}{2}\Bigl \{ d\sigma(\pi^+ p)+ 
d\sigma(\pi^- p)- d\sigma(\pi^0 n\to \pi^- p)\Bigr \},
\end{equation}
 where $ d\sigma(\pi^+ p)$, $d\sigma(\pi^- p)$ and 
$d\sigma(\pi^0 n\to \pi^- p)$ 
are the differential cross section of the $\pi^+p$ and
$\pi^- p$ elastic scattering and  charge exchange reaction 
$\pi^0 n\to \pi^- p$, respectively.
After the amplitude $a_{\frac{3}{2}}$ is excluded from  Eq. (\ref{api0}),
the cross section of the $\pi^0 p$ scattering  takes the form  
\begin{equation}
\label{pi0psecm}
 d{\widetilde \sigma}(\pi^0 p\to \pi^0 p)=
\frac{1}{18}\Bigl \{ 3d\sigma(\pi^- p)- d\sigma(\pi^+ p)+ 
3d\sigma(\pi^0 n\to \pi^- p)\Bigr \}.
\end{equation}
 In order to exclude the term $a_\frac{3}{2}$  from the $\pi^0 p$ elastic
 scattering in calculation of the cross section of the reaction
 $pp\to \{pp\}_s\pi^0$
 one should substitute  Eq. (\ref{pi0psecm}) instead of
 Eq. (\ref{pi0psec})
 into Eq. (\ref{pppppi0}).
 When we do so, the absolute value of the calculated cross section of the
 reaction $pp\to \{pp\}_s\pi^0$ at 0.4 - 1.1 GeV diminishes by two orders
 of magnitude and comes
 in  strong contradiction with the data (see dashed line in
 Since the 
amplitude $a_{\frac{3}{2}}$
 of the $\pi N$ elastic 
scattering is dominated by the $\Delta(1232)-$isobar at 
$\sqrt{s_{\pi N}}\sim 1.15-1.35$ GeV,
 this analysis  shows that
 the excitation of the
 $\Delta(1232)$-isobar 
 dominates in the reaction $pp\to \{pp\}_s\pi^0$ at 0.5 - 1.0 GeV too.
 On the other hand,  since the $s-$wave 
 intermediate $\Delta-N$ state is forbidden,
 but was not excluded  from the reaction amplitude
 within this model, the 
 agreement obtained between the calculated and measured shape of the cross
 section suggests that the 
absence  of this s-state in an exact OPE amplitude would be 
not so crucial for 
the reaction $pp\to \{pp\}_s\pi^0$ at 0.4-1.1 GeV,
 as might follow from the qualitative arguments given in the Introduction.
 In other words, it would mean  that the p-wave and higher odd waves are not 
 suppressed drastically by centrifugal barrier (perhaps, due to long-range
 character of the $\Delta-N$ interaction) and make a large enough 
 contribution to this reaction.

  Let us make some further comments.
  Firstly, the second maximum of the forward $pp\to \{pp\}_s\pi^0$
  cross section is observed at 1.97 GeV. The forward 
 $pp\to d\pi^+$ cross section exhibits a similar maximum 
  \cite{arndt}.
 This peculiarity of the $pp\to d\pi^+$ cross section
 was interpreted in Ref. \cite{yao} within the OPE model
 as  a manifestation of  heavy nucleon resonances in the elastic 
 $\pi N$ scattering.
 One can see from Fig.\ref{fig2} that the OPE model considerably
 underestimates the
 magnitude of the observed second maximum in the $pp\to \{pp\}_s\pi^0$ 
 cross section.
One may suppose that excitation of heavy $\Delta$'s is not sufficient
to explain the data at 2 GeV and, therefore, other  mechanisms of this
reaction like $N^*$ exchange or Reggeon exchange recently discussed
 in Ref.\cite{uzhw2007} make  a  sizeable contribution in this region.
 To  choose  between the heavy $\Delta-$isobars excitation  and  the
 $N^*$ (or Reggeon) exchange mechanism one should measure  the ratio
 of the cross sections $pp\to \{pp\}_s \pi^0$ and $pn\to \{pp\}_s\pi^-$
\cite{uzhw2007}.
 
 Secondly, as can be shown, the present model predicts a
 smooth increase (5-15\%)
 of the differential cross  section in forward direction at
 $\theta= 0^\circ-15^\circ$, that is in qualitative agreement with the data
 at 2 GeV, but in disagreement  at lower energies. 
 A more detailed model, with distortions and explicit
 $\Delta$-isobars included, has to be developed  
 to describe the angular dependence of the cross section. 
 This kind of model considered in  Ref. \cite{niskanen06} below 0.9 GeV
 was to some extent successful in this  respect, while failed to
 describe the  energy dependence.    

\section{Conclusion}

    Arguments, based on parity and angular momentum conservation,
  show that the S-wave $\Delta$ N-intermediate state is forbidden
  in the reaction $pp\to \{pp\}_s\pi^0$, when the final pp-pair is produced
  in the $^1S_0$-state. The microscopical model \cite{niskanen06},
 which takes into account this specific feature of the reaction
 $pp\to \{pp\}_s\pi^0$, since  includes explicitly the $\Delta$-isobar
 contribution via coupled $NN-$ and $N\Delta-$channels, is in strong  
 disagreement with the observed energy dependence of the recently measured
 cross section of the reaction $pp\to \{pp\}_s\pi^0$ at 0.5 - 1 GeV.
 On the other hand,  a rather simple  OPE model
 developed in the present work,
 which includes the
 subprocess $\pi^0 p\to \pi^0 p$ and the final state $pp(^1S_0)$-interaction, 
  reproduces the observed shape of energy dependence
 of the cross section of the reaction $pp\to \{pp\}_s\pi^0$ at 0.5 - 1.4 GeV
 and to some extent agrees with its absolute value. Thus, the OPE model
 clearly exhibits
 dominance of the $\Delta(1232)$-isobar in this region, although the angular
 dependence of the $pp\to \{pp\}_s\pi^0$ cross section is not described within
 this simple model.
 One should note, that a quite similar OPE model was recently successfully
 applied to the $pd\to \{pp\}_sn$ reaction in Ref.\cite{uzhw2007} just in the 
 $\Delta$-isobar region. Therefore,
 a failure of the model of Ref. \cite{niskanen06}, most likely, is related not
 to the $\Delta$ contribution itself, but rather caused by
 interference effects with
 other terms, for example, with phenomenological heavy meson exchange. 
 More insight into the dynamics of the single pion production in pN collision
 can be gained by further measurement of the reaction $pn\to \{pp\}_s\pi^-$
 at the same kinematical conditions. It would be also interesting to get data
 on the reaction $pp\to \{pp\}_s\pi^0$ at higher excitation energy of the
 final pp-pair, $E_{pp}=3-10$ MeV, where small components of the pp-wave
 function start to contribute and allow the S-wave intermediate $\Delta N-$
 state.

  I am grateful to V.I.~Komarov, A.V.~Kulikov and V.~Kurbatov for useful
 comments.

\end{document}